\documentstyle[psfig,conf_iap,]{article}
\begin{document}
\heading{%
%
3D Spatial Distribution of the Intergalactic Medium
\\
%
} 
\par\medskip\noindent
\author{%
Patrick Petitjean$^{1,2}$, Emmanuel Rollinde$^{1}$, Bastien Aracil$^{1}$,
Christophe Pichon$^{1,3,4}$, St\'ephane Colombi$^{1,4}$

}
\address{%
Institut d'Astrophysique de Paris, CNRS, 98bis Boulevard Arago, 
       F--75014 Paris}
\address{%
UA CNRS 173-DAEC, Observatoire de Paris-Meudon, F-92195 Meudon Cedex}
\address{%
Observatoire de Strasbourg, 11 rue de l'Universit\'e, 67000 Strasbourg,
France}
\address{%
Numerical Investigations in Cosmology (N.I.C.), CNRS, France}

\begin{abstract}
Very recently a new inversion method has been developped to analyze
the intergalactic medium seen in absorption in quasar spectra
(the so-called Lyman-$\alpha$ forest). This method is applied
to recover the temperature of the gas and the underlying density field.
Using constraints from the Lyman-$\beta$ forest, it is possible to
recover this field up to over-densities $\delta$~$\sim$~10. 
By inverting the H~{\sc i} and C~{\sc iv} absorptions together it has been
shown that the C~{\sc iv}/H~{\sc i} ratio varies through the 
profile of strong lines, beeing larger in the wings.
The method can be applied to reconstruct the 3D density field from
multiple lines of sight and is shown to give good results
up to mean separations of 3~arcmin. Results from a survey of QSO pairs
performed with HST/STIS and VLT/UVES-FORS are summarized. 
\end{abstract}
\section{Introduction}
The numerous absorption lines seen in the spectra of distant quasars
(the so-called Lyman-$\alpha$ forest) reveal the
intergalactic medium (IGM) up to redshifts larger than 5.
It is believed that the space distribution of
the gas traces the potential wells of the dark matter.
Indeed, recent numerical $N$-body simulations have been successful at
reproducing the observed characteristics of the Lyman-$\alpha$ forest
(e.g. \cite{cen}\cite{petitjean1}\cite{hernquist}).
The IGM is therefore seen as a smooth all-pervading medium which can be 
used to study
the spatial distribution of the mass on scales larger than the Jeans' length.
This idea is reinforced by observations of multiple lines of sight.
It is observed that the Lyman-$\alpha$ forest is fairly homogeneous
on scale smaller than 100~kpc (e.g. \cite{smette}) and highly 
correlated on scale up to one megaparsec (e.g. \cite{petitjean2}\cite{crotts}).
 The number of suitable multiple lines of sight is small 
however and the sample need to be significantly enlarged before any firm 
conclusion can be drawn (see Section 3).
\par\noindent
Very recently, new methods have been implemented to recover the real density 
space distribution of the IGM by inversion of the Lyman-$\alpha$ forest 
\cite{nusser}\cite{pichon}. It is possible to derive the 
physical state of the gas (temperature, density, metallicity) 
by inversion of high S/N ratio and high resolution spectra taken with 
UVES (see Section 2). The method 
has also been tested to recover the 3D topology of the large scale structures 
when applied to the inversion of a network of adjacent lines of sight 
(see Section 3).
\par\noindent
\section{Inversion of the Lyman-$\alpha$ forest}
\begin{figure}
\centerline{\vbox{
\psfig{figure=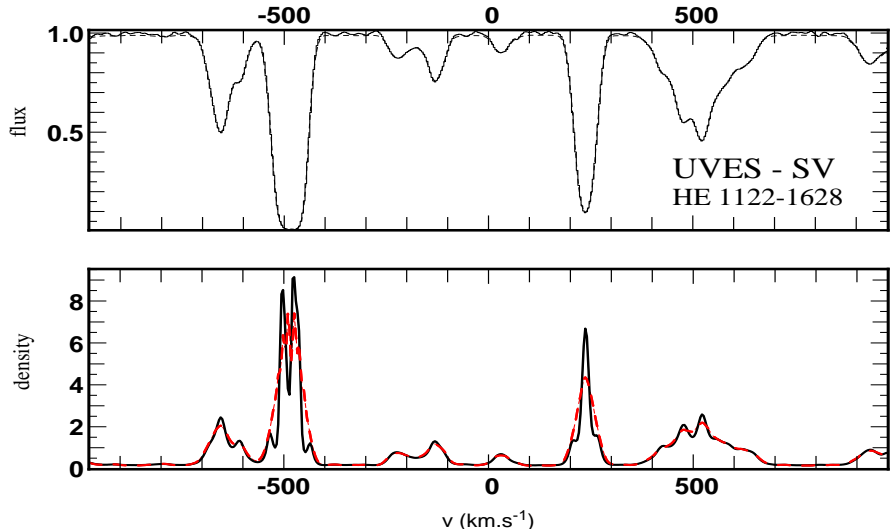,height=7.cm,width=12.cm}
}}
\caption[]{Part of the high resolution and high S/N spectrum of the 
QSO HE 1122-1628. The normalised flux is fitted (upper panel, the 
 solid line is the data and  the dashed line the fit model)
and the corresponding density field is recovered (lower panel).
The broadening of the recovered density peaks depends on the temperature of
the IGM. High temperature yields narrow  profiles (solid line).
Lower temperature yields  smoother  profiles (dashed line).
}
\label{f:HE1122}
\end{figure}

We have developped a new inversion method to recover the one dimensional 
density field from the fit of normalised QSO spectra\cite{pichon}.
An example of this reconstruction is given in Fig.~\ref{f:HE1122}
which shows part of the spectrum of HE 1122-1628 (upper panel) and
the reconstructed density (lower panel). The fit is overplotted in the upper 
panel as a dashed line.\\
To illustrate the dependence of the results on the gas temperature, 
the density field is reconstructed  for different assumed constant 
temperatures. 
High temperatures yield narrow reconstructed density profiles 
(solid line in the lower panel of Fig.~\ref{f:HE1122}); lower temperatures
yield smoother profiles (dashed line). Therefore, although some degeneracy
exists (see \cite{rollinde}), one can constrain the 
temperature at mean density of the IGM and its cosmological evolution. 
\par\noindent
Using the high resolution, high S/N spectrum of QSO HE1122-1628
obtained during Science Verification of UVES we derive
a low temperature of $10^4$ K at $z$~$\sim$~2.2.
A more detailled description of this procedure and
other applications of the method may be found in \cite{pichon,rollinde}.
In particular, using additional constraints from the Lyman-$\beta$ forest, 
it is possible to recover the density field up to over-densities 
$\delta$~$\sim$~10. In addition by inverting the H~{\sc i} and C~{\sc iv} 
absorptions together it has been shown that the C~{\sc iv}/H~{\sc i} 
ratio varies through the profile of strong lines, being larger in the wings
\cite{rollinde}.
%
\begin{figure}
\centerline{\vbox{
\psfig{figure=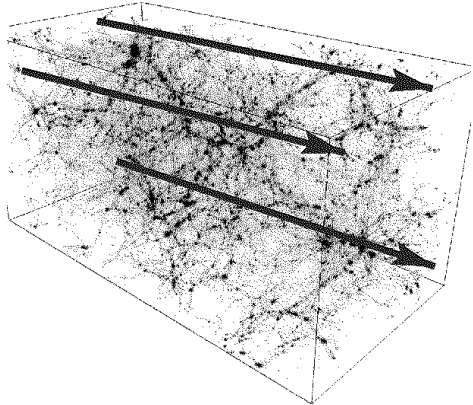,height=8.cm}
}}
\caption[]{Dark matter distribution in the CDM simulation, at z=2.
 Darker regions are denser (logarithmic scale).
 A grid of lines of sight through the box is defined along  
 the direction shown by the arrows. The inversion method is then applied
to the synthetic spectra and results are compared to the simulation
(see Fig. 3).
}
\label{f:simulation}
\end{figure}
\section{Large scale structures }
We have applied the method  to recover the 3D spatial
distribution of the IGM using multiple lines of sight \cite{pichon}.
It must be emphasized that the spectral resolution along
each line of sight should be 
of the order or slightly higher than the corresponding 
spatial resolution defined as the mean separation between lines of sight.
This means that intermediate-resolution of the kind achieved by VLT/FORS 
is enough for the corresponding observational programme provided 
the blue sensitivity is large enough... of course !
%
\par\noindent The method has first been tested  on simulations. 
For this we use $N$-body SCDM simulations performed at the
Institut d'Astrophysique de Paris.

\begin{figure}
\centerline{\vbox{
\psfig{figure=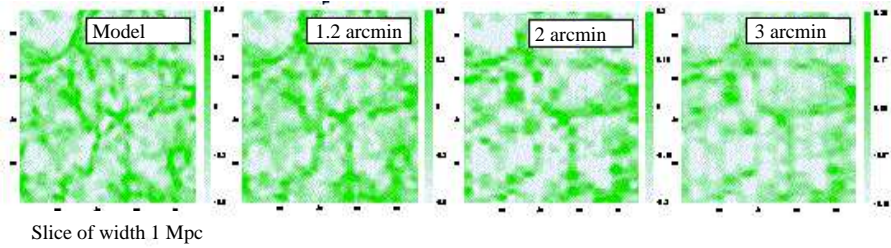,height=3.3cm}
}}
\caption[]{
Left panel: Spatial distribution of the DM in a slice of 1 Mpc taken through 
the simulation (Fig. 2). Other panels : reconstructed distribution 
using a grid of lines of sight separated by, respectively, 
1.2, 2 and 3 arcmin.}
\label{f:reconstruction}
\end{figure}
\begin{figure}
\centerline{\vbox{
\psfig{figure=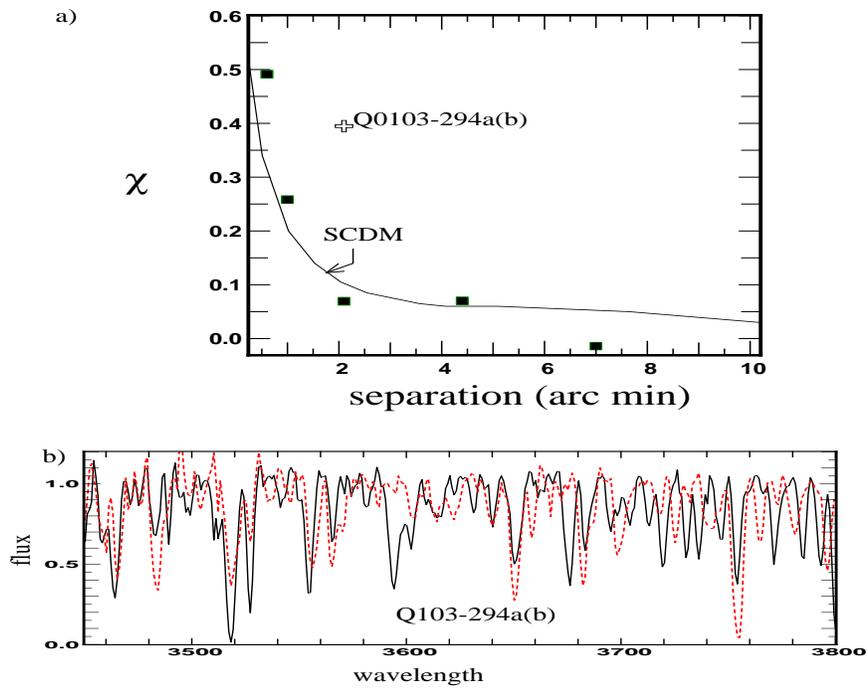,height=10.4cm,width=11.5cm}
}}
\caption[]{Transverse correlation coefficient $\chi$ plotted as a 
function of angular separation (panel a). Squares are the observed numbers
from FORS spectra of QSO pairs. The solid curve is the average value from
hundred pairs taken in the SCDM simulation, for each separation.
The pair Q~0103-294A,B exhibits a strong correlation. This 
is apparent on the spectra shown in the bottom panel.} 
\label{f:chi}
\end{figure}
\par\noindent We extract a grid of lines of sight\  through  
the simulation box (see Fig.~\ref{f:simulation}) and obtain the corresponding  
absorption spectra. The inversion
method is then used to recover the initial 3D density field from
the 1D information along these lines of sight.
Results are shown in Fig.~\ref{f:reconstruction} for a slice of 
width 1~Mpc (for illustrative purpose, we use 
1 Mpc $\simeq\  1h$ arcmin; $h=H_0/(100$ km/s/Mpc)) 
taken through the simulation box. The left panel shows the spatial
distribution of the dark matter as derived from the $N$-body simulation. The
other panels show the reconstructed distributions using a grid of
lines of sight separated by, respectively, 1.2, 2 and 3~arcmin.
As expected, structures are better recovered for smaller separations.
However, even with 3 arcmin separation, it is still possible to recover the 
topology of the large scale structures.
\par\noindent To reach this angular separation,  a density of 
QSOs larger than 100 per square degree is needed. This
 implies to observe QSOs of 
magnitude of the order of  21.5 or fainter (see \cite{petitjean3})
{\sl and } to chose the
best field (to be found) where the QSO density is highest.
Large surveys of QSOs are underway and such
a field will probably be available very soon.
%
%
\par\noindent In the mean time, we have started a study of the correlation 
between pairs of QSOs. 
From the transverse  correlation it is possible to derive 3D information on
the density field \cite{hui2}\cite{mcdonald}.
Here we test the transverse correlation with a convenient single number,
$\chi$\cite{viel}:
\begin{equation}
\chi(r_{\perp})=\int_{Ly\alpha} \frac{({\cal F}_0-\overline{{\cal F}_0})({\cal F}_1-\overline{{\cal F}_1})}{\sigma_{{\cal F}_0}\sigma_{{\cal F}_1}}\,,
\label{e:chi}
\end{equation}
where ${\cal F}_0$ and ${\cal F}_1$ are the two normalised spectra
of a pair of quasars separated by a transverse distance $r_{\perp}$.
The value of $\chi$ gives a quantitative estimate of the correlation,
ranging from 0 if the spectra are uncorrelated to 1 for identical spectra.
We derive $\chi$ versus $r_{\perp}$ in the simulation 
(solid curve in Fig.~\ref{f:chi}). This curve is model dependent
mainly trough the relation between angular and comoving distances, 
but also through the size of the structures, etc...
This method yields therefore an additional cosmological test.
About twenty pairs are needed for each separation bin to discriminate 
between different cosmologic models\cite{viel}.
\par\noindent
We measure $\chi$ on our recent FORS observations of 10 QSO pairs
at $z$~$\sim$~2.2 (filled squares on Fig.~4).
The observational points are consistent with the SCDM model. 
However, an interesting pair shows a much stronger correlation. 
This strong correlation is apparent from the spectra (bottom panel
in Fig.~4) and has no
explanation yet. It is also important to realise that the correlation is 
still significant at 4 arcmin. 
\par\noindent 
%
At smaller redshift, $z$~$\sim$~1, new HST data \cite{impey,aracil} show that
a correlation signal is present up to 3~arcmin (1.5$h^{-1}_{50}$~Mpc physical).
However the coefficient $\chi$ is smaller than at high redshift 
flagging expected cosmological
evolution. The later is probably mostly due to evolution in the strength of  
the absorption lines. Comparison are difficult however because of small 
statistics but also and mostly because of comparatively smaller sensitivity 
of HST observations. Future observations with new instruments will overcome 
these difficulties.

\begin{figure}
\centerline{\vbox{
\psfig{figure=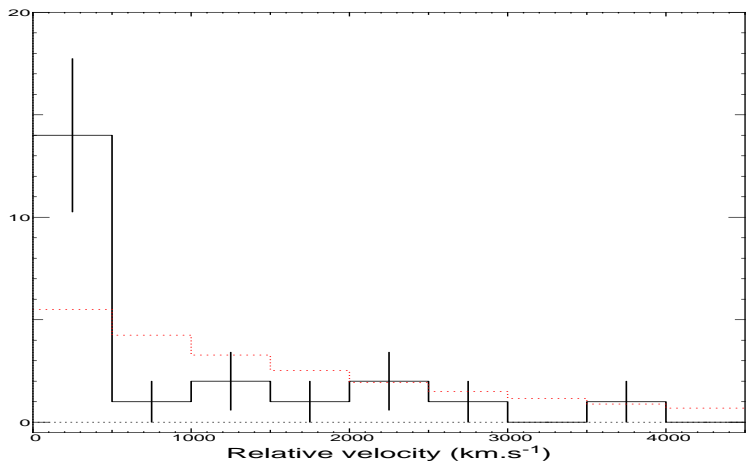,height=6.cm,width=10.cm}
}}
\caption[]{Number of matches between $z$~$\sim$~1 absorptions seen along 
the two lines of sight of quasar pairs with separations between 2 and 3 arcmin 
($\sim$ 1 to 1.5~$h^{-1}_{50}$~Mpc physical) 
against the velocity difference between the two absorptions. The dashed line
corresponds to the expected numbers for randomly distributed lines.
It is apparent that there is an excess of matches for $\Delta
V$~$<$~350~km~s$^{-1}$.
} 
\label{f:hst}
\end{figure}
%
%


\acknowledgements{This work is based on
observations carried out at the European Southern Observatory (ESO) 
with the UVES and FORS spectrographs on the Very Large
Telescope (VLT) at the Cerro Paranal Observatory in Chile and on observations
with the NASA/ESA Hubble Space Telescope, obtained at STScI, which is
operated by the AURA, under NASA contract NAS5-26555.
}

\vfill
\end{document}